\documentclass[aps,twocolumn,linenumbers]{revtex4}

\usepackage[english]{babel}

\usepackage[letterpaper,top=2cm,bottom=2cm,left=2cm,right=2cm,marginparwidth=1.75cm]{geometry}

\usepackage{amsmath}
\usepackage{amsfonts}
\usepackage{graphicx}
\usepackage{tabularx}
\usepackage{array}
\usepackage{calc}

\usepackage[colorlinks=true, allcolors=blue]{hyperref}

\begin{document}

\title{Multi-dimensional frequency-bin entanglement-based quantum key distribution network}

\author{George Claudiu Crisan$^1$, Antoine Henry$^{1,2}$, Dario A. Fioretto$^1$, Juan R. Alvarez$^{1,3}$, Stéphan Monfray$^4$, Frédéric Boeuf$^4$, Laurent Vivien$^1$, Eric Cassan$^1$, Carlos Alonso-Ramos$^1$, Nadia Belabas$^1$}
\affiliation{$^1$Centre de Nanosciences et de Nanotechnologie, CNRS , Université Paris Saclay, Palaiseau, France\\
$^2$Department of Mechanical Engineering and Institute for Physical Science and Technology, University of Maryland, College Park, MD 20742, USA\\
$^3$LTCI, Telecom Paris, Institut Polytechnique de Paris, Palaiseau, France\\
$^4$ST Micro Electronics SAS, France\\
}

\begin{abstract}
    Quantum networks enhance quantum communication schemes and link multiple users over large areas. Harnessing high dimensional quantum states - i.e. qu-d-its - allows for a denser transfer of information with increased robustness to noise compared to qubits.
    Frequency encoding enables access to such qu-d-its at telecom wavelengths, while manipulating quantum information with off-the-shelf fibered devices.
    
    We use a low free spectral range silicon microresonator to generate Bell states of dimension $d$=2 (qubits) and $d$=3 (qutrits) via spontaneous four wave mixing, to implement and optimize a multi-dimensional frequency-bin entanglement-based quantum key distribution network.
    
    We tune the source (via pump power), the signal processing (via coincidence window size) and qu-$d$-it encoding ($d$=2 or 3) using a single fibered hardware based on Fourier-transform pulse shaping and electro-optic modulation depending on interconnection lengths. We achieve secure key rates of $1374$ bit/s with qutrits, and estimate the communication range to $295$ km with qubits.
    We access up to $80$ frequency modes, resulting in $21$ quantum channels, that provide stable communication over more than $21$ hours.  
    We demonstrate a competitive communication range in a multi-dimensional entanglement-based quantum key distribution network that lays the groundwork for larger dimensionality implementations deployed on metropolitan fiber links.
\end{abstract}

\maketitle

\section{Introduction}

Quantum networks are desirable architectures in which quantum resources in transferred securely between distant nodes, for example quantum processors, quantum sensors or classical users.
Quantum Key Distribution (QKD) allows physically secure communications between two distant users.
The number of users and the communication range can be extended with trusted nodes \cite{chen_integrated_2021}, but expose the quantum network to security risks.
Entanglement based QKD protocols such as BBM92 \cite{bennett_quantum_1992} are compatible with networks and do not require the use of trusted nodes \cite{fanyuan_robust_2022}. 
In addition, entanglement-based protocols provide a better distance scaling factor and security against general attacks when compared to prepare-and-measure protocols \cite{ma_quantum_2007}.

While polarization encoding is widely used in high performance QKD links using simple and cost-effective optical elements \cite{appas_flexible_2021,li_high-rate_2023,liu_experimental_2019,yang_high-rate_2024}, it is prone to polarization shifts and can only implement two dimensional quantum states.
In contrast, high dimensional quantum states, qu$d$its, offer greater information capacity and noise resilience than qubits \cite{cerf_security_2002}, enabling more efficient quantum algorithms \cite{cozzolino_orbital_2019} and enhancing quantum communication protocols \cite{wang_qudits_2020} by improving communication rates and raising the noise tolerance threshold \cite{cerf_security_2002}.

Time-bin, Orbital Angular Momentum (OAM) and frequency-bin encoding enable easy access to high dimensional quantum states. 
Frequency encoding can be used for QKD without the need for complex interferometers for state measurements, as required in time-bin encoding \cite{islam_provably_2017,lee_large-alphabet_2019,Steiner2023,yu_quantum_2025}.
Additionally, frequency encoding offers practical benefits like dense single-mode fiber transmission, which OAM encoding cannot accommodate \cite{sit_high-dimensional_2017,bouchard_experimental_2018,cozzolino_high-dimensional_2019}, as well as resilience against polarization instabilities and parallelizable operations, making it suitable for quantum network applications.

Pioneer frequency encoding QKD protocols have used side-bands or sub-carrier generation with attenuated coherent laser pulses \cite{merolla_single-photon_1999,merolla_integrated_2002,silva_optical_2008,pinheiro_one-way_2011,mora_experimental_2012,gleim_secure_2016}, where the logical $|0\rangle$ and $|1\rangle$ states are differentiated using the phase carried by the side bands \cite{merolla_integrated_2002}. 
The states encoded with the sub-carrier phase method have been shown to be incompatible with entanglement based protocols, and are currently lacking a security proof against general attacks \cite{sajeed_approach_2021}.
In contrast, frequency-bin encoding uses the frequency degree of freedom to encode logical quantum states on single photons.
The frequency-bin encoding is compatible with high $d$-dimensional quantum states, qu$d$its, that have been demonstrated up to $d = 8$ \cite{lu_bayesian_2022}, while also retaining the frequency multiplexing capabilities developed for classical telecommunications.
Frequency-bin qu$d$its can be manipulated at telecom wavelengths through the use of commercial Electro-Optic-Modulators (EOMs) and Programmable Filters (PFs) \cite{kues_-chip_2017,lukens_frequency-encoded_2017,lu_electro-optic_2018} for parallel and independent quantum gates \cite{henry_parallelizable_2023}.

Frequency-bin entangled states can be generated on high-index glass microresonators (MR) \cite{kues_-chip_2017}, lithium niobate waveguides \cite{henry_correlated_2023,lu_bayesian_2022,imany_50-ghz-spaced_2018,lu_electro-optic_2018,cabrejo-ponce_high-dimensional_2023,henry_generation_2023,lu_quantum_2019}, AlGaAs MRs \cite{pang_versatile_2025}, SiN MRs \cite{imany_50-ghz-spaced_2018,lu_bayesian_2022,mahmudlu_fully_2023}, and Silicon-On-Insulator MRs \cite{henry_nm_2024,clementi_programmable_2022,borghi_reconfigurable_2023}.
Silicon CMOS (complementary metal-oxyde-semiconductor) compatible technology offers high scalability and high-quality-factor manufacturing.
The silicon platform also provides key advantages for the generation and manipulation of frequency-bin encoded photon pairs since it has a high $\chi^{(3)}$ non linear component ($n_2=5.10^{-18}\text{m}^{-2}\text{W}^{-1}$) compared to SiN ($n_2=3.10^{-19}\text{m}^{-2}\text{W}^{-1}$) at $1550$ nm, enabling the generation of biphotonic frequency combs through Spontaneous Four Wave Mixing (SFWM), while operating at room temperature.

In this work we demonstrate a proof of principle of a frequency-bin encoded BBM92 entanglement-based QKD network with $d=2$ and $d=3$ qu$d$its, i.e. qutrits, harnessing reconfigurability to address specific user needs.
This work builds on an entanglement-based frequency-bin QKD demonstration with qubits \cite{henry_nm_2024}, with only two other groups having recently implemented frequency-bin encoded QKD protocols with qubits \cite{khodadad_kashi_frequency-bin-encoded_2025,tagliavacche_frequency-bin_2025}.
In \cite{khodadad_kashi_frequency-bin-encoded_2025} a frequency-bin encoding BBM92 QKD network with 3 channels is demonstrated. A secure communication rate of $9$ bit/s and an error rate of $7\%$ are achieved at $0$ dB applied attenuation on the quantum channel. The communication is secure up to a maximum simulated attenuation of $51.5$ dB.
In \cite{tagliavacche_frequency-bin_2025} a frequency-bin encoded BBM92 protocol is demonstrated though a fiber spool. A communication rate of $110$ bit/s is achieved at $0$ km, with an error rate averaging $10\%$. The communication range is estimated at $30$ km and phase stabilization techniques are employed to adress environmental fluctuations in optical fibers.

We leverage the experimental conditions of the source and the information processing to increase the secure communication rate and the communication range of a QKD network employing both $d=3$ and $d=2$ qu$d$its.
In particular, we find that the optimal pump power and coincidence window for qutrits are lower than those for qubit implementations.
We achieve an averaged secure rate above $1$ kbit/s with qutrits and above $450$ bit/s for qubits across 21 QKD channels, at $0$ dB applied attenuation. The error rates are below $12\%$ for qutrits and below $8\%$ for qubits, guaranteeing the security of the $21$ channels. The communication range is estimated to $295$ km ($59$ dB of total attenuation) with qubits.
This work paves the way for the development of scalable and reconfigurable high dimensional frequency-bin quantum networks.

\begin{figure*}
    \centering
    \includegraphics[width=\linewidth]{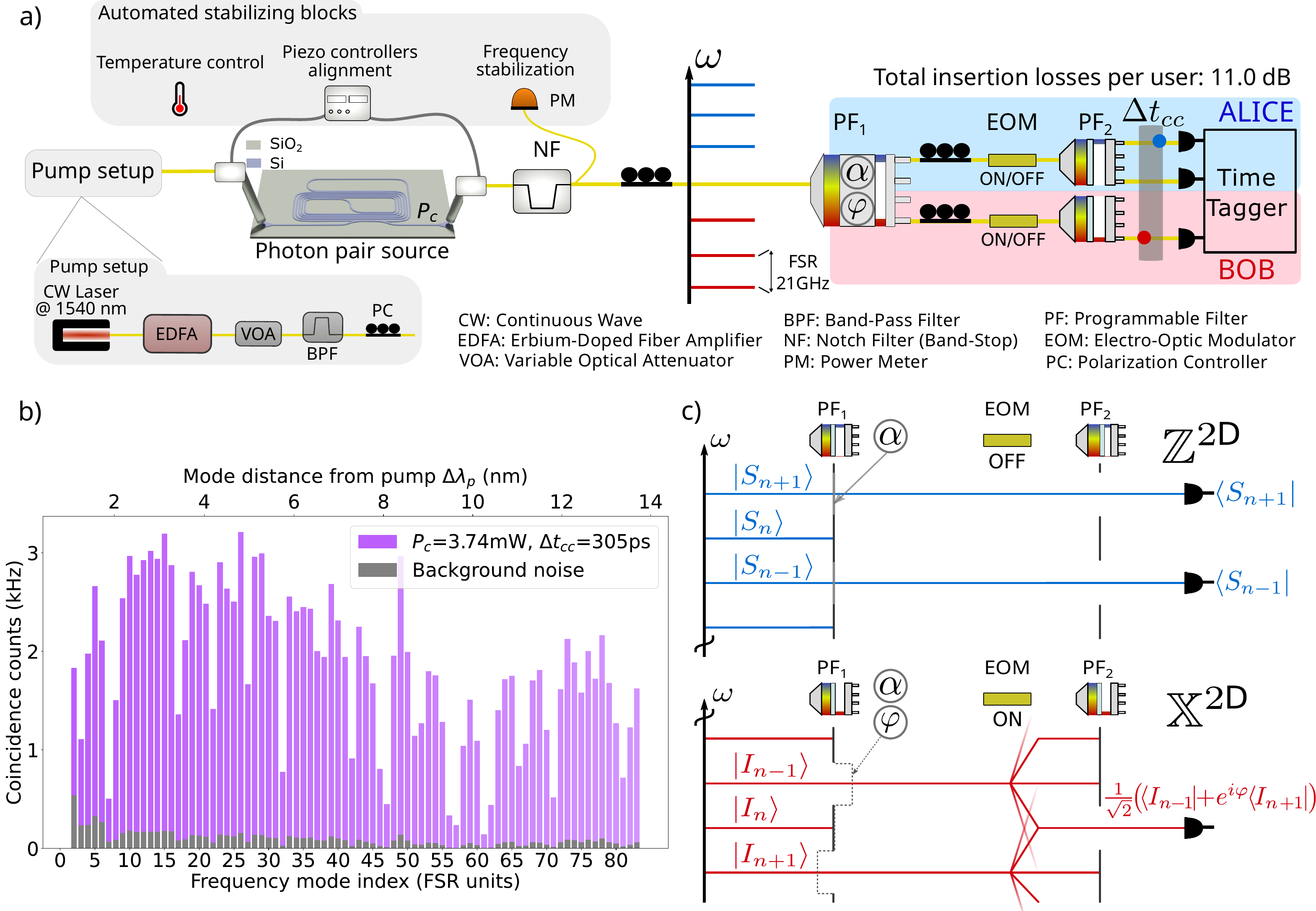}
    \caption{\textbf{a)} Simplified experimental setup for the multi-dimensional frequency-bin encoded BBM92 protocol implementation.
    The pump and collection fibers can be automatically aligned.
    The pump wavelength is actively adjusted to the resonance at $ \omega_p = 1539.970$ nm of the silicon resonator. 
    Details are given in Methods section \ref{methods}.
    The signal (blue) and idler (red) entangled photons of the frequency comb generated by Spontaneous Four Wave Mixing (SFWM) are distributed to Alice and Bob respectively. 
    The attenuation $\alpha$ and the phase $\varphi$ on each frequency mode are applied by PF$_1$.
    ON/OFF indicates the status of the EOMs.
    The coincidence window $\Delta t_{cc}$ represents the temporal interval within which detected events count as coincidences.
    \textbf{b)} Joint Spectral Intensity (JSI) of the photon pair source, including coincidence count rates for signal and idler pair indexed by the spectral separation from $\omega_p$, in Free Spectral Range (FSR) units.
    The JSI measurement was performed at a power on chip $P_c=3.74$ mW and $\Delta t_{cc}=305$ ps.
    \textbf{c)}
    Example sketch of the frequency-bin encoded BBM92 protocol measurements in $\mathbb{Z}^{2D}$ and $\mathbb{X}^{2D}$ basis for qubits.
    The quantum channel is $3$ resonance-wide ($63$ GHz-wide).
    The attenuation $\alpha$ simulates the distance related-losses between Alice and Bob.
    In the $\mathbb{X}$ basis, PF$_1$ controls the state projections via $\varphi$ and the EOM mixes the two frequency modes on a common frequency channel singled-out by PF$_2$. Details about the qutrit implementation are given in the main text.}
    \label{fig:Setup}
\end{figure*}

\section{Experimental BBM92 protocol implementation}

In Fig.\ref{fig:Setup}a we show a simplified version of the experimental setup. More details about the experimental setup can be found in the Methods section \ref{methods}.
We continuously pump a microresonator using a tunable laser that we amplify and attenuate to control the power on chip $P_c$ from $0$ to $6$ mW.

Our source is a spiral resonator etched from a $300$ nm silicon layer on top of a $3$ $\mu$m buried oxyde layer, compatible with silicon microelectronics industry.
The resonator is coupled to a bus waveguide along a coupling length of $10$ nm. The input and output flat cleaved fibers, set at $15$ degree incidence angle, access the bus waveguide via surface grating coupler on both ends of the bus waveguide.
Our source is a $3.54$ mm long silicon spiral resonator that is folded into a minimal footprint of $165$ $\mu$m by $225$ $\mu$m.
The spiral shape reduces strain and substrate instabilities, enabling a Free Spectral Rage (FSR) of $21.23$ GHz, that is constant to $2\%$ (less than $0.45$ GHz), for $88\%$ of frequency modes from $1526$ to $1554$ nm.
The 700 nm-wide waveguide ensures phase-matching conditions for Spontaneous Four Wave Mixing (SFWM) by reducing the anomalous dispersion \cite{medina_these}, and enables the efficient generation of spectrally entangled photon pairs. 
The photon pair consists of a signal and an idler photon generated in a symmetric frequency comb with respect to the frequency of the pump $\omega_p= 194.67$ THz ($1539.970$ nm). 
Additionally, the high Q factor of $4.75 \cdot 10^{5}$ of the cavity makes for a state-of-art brightness source of $5.1 \pm 3$ pairs.s$^{-1}$mW$^{-2}$GHz$^{-1}$ at this FSR \cite{henry_nm_2024}.

The frequency modes generated by the resonator are indexed by $n \in \mathbb{N}$: $I_n = \omega_p - n\cdot \text{FSR}$ for idler mode and $S_n = \omega_p + n\cdot \text{FSR}$ for signal mode.
The quantum state $|\Psi\rangle$ of the photon pair emitted within the frequency comb is given by

\begin{equation}
\label{eqn:Frequency}
    |\Psi\rangle=\frac{1}{\sqrt{N}}\sum_{n=1}^Ne^{i\phi_n}|I_n\rangle|S_n\rangle,
\end{equation}
where $N$ is the number of available frequency modes limited by the $5$ THz bandwidth of a Programmable Filter (PF), and $\phi_n$ is the residual spectral phase \cite{lu_bayesian_2022} which varies as $\phi_{n+1} - \phi_{n} = 0.065$ rad for our sample.

The two photons are spatially separated and projected onto the biphoton state $\langle S_m I_m | \Psi\rangle$ by the spectral filtering of the PFs, where $m$ is the resonance frequency mode index ranging from $3$ to $83$ in FSR units.
The resulting coincidence count rates measured by single photon detectors are given by the Joint Spectral Intensity (JSI) shown in Fig. \ref{fig:Setup}b, in purple, while the background counts are shown in gray. 
 
The natural frequency entanglement of the photon pair enables the generation of Bell states, used in entanglement-based QKD protocols.
In the BBM92 protocol \cite{bennett_quantum_1992}, a biphotonic Bell State is distributed to two parties, Alice and Bob. 
Each party measures the received photon in one of two randomly selected Mutually Unbiased Bases (MUBs). 
Following a sufficient number of measurement rounds and subsequent basis reconciliation, the quantum correlations between their measurement outcomes enable the generation of a shared secure key, used to encode secure messages.
By leveraging the $83$-resonance bandwidth of our photon source and the access to high dimensional frequency-bin encoded Bell states, we employ frequency multiplexing to support secure communication among multiple users simultaneously.

We use PF$_1$ of Fig. \ref{fig:Setup}a to spatially separate the signal and idler photons and distribute them to two users of a quantum channel. 
The quantum channel is $63$ GHz-wide ($3$ resonance-wide) and contains either three frequency modes for the generation of a $d=3$ qu$d$it (qutrit) Bell state, noted as
\begin{equation}
\label{eqn:Psi3}
    |\Psi\rangle_n^{3 \text{D}} = \frac{1}{\sqrt{3}} \left(  |I_{n-1}S_{n-1}\rangle + |I_{n}S_{n}\rangle + |I_{n+1}S_{n+1}\rangle \right)
\end{equation}
or two frequency modes for the creation of a $d=2$ qu$d$it (qubit) Bell state noted as 
\begin{equation}
\label{eqn:Psi2}
    |\Psi\rangle_n^{2 \text{D}} = \frac{1}{\sqrt{2}} \left(  |I_{n-1}S_{n-1}\rangle + |I_{n+1}S_{n+1}\rangle \right),
\end{equation}
\begin{figure}
    \centering
    \includegraphics[width=\linewidth]{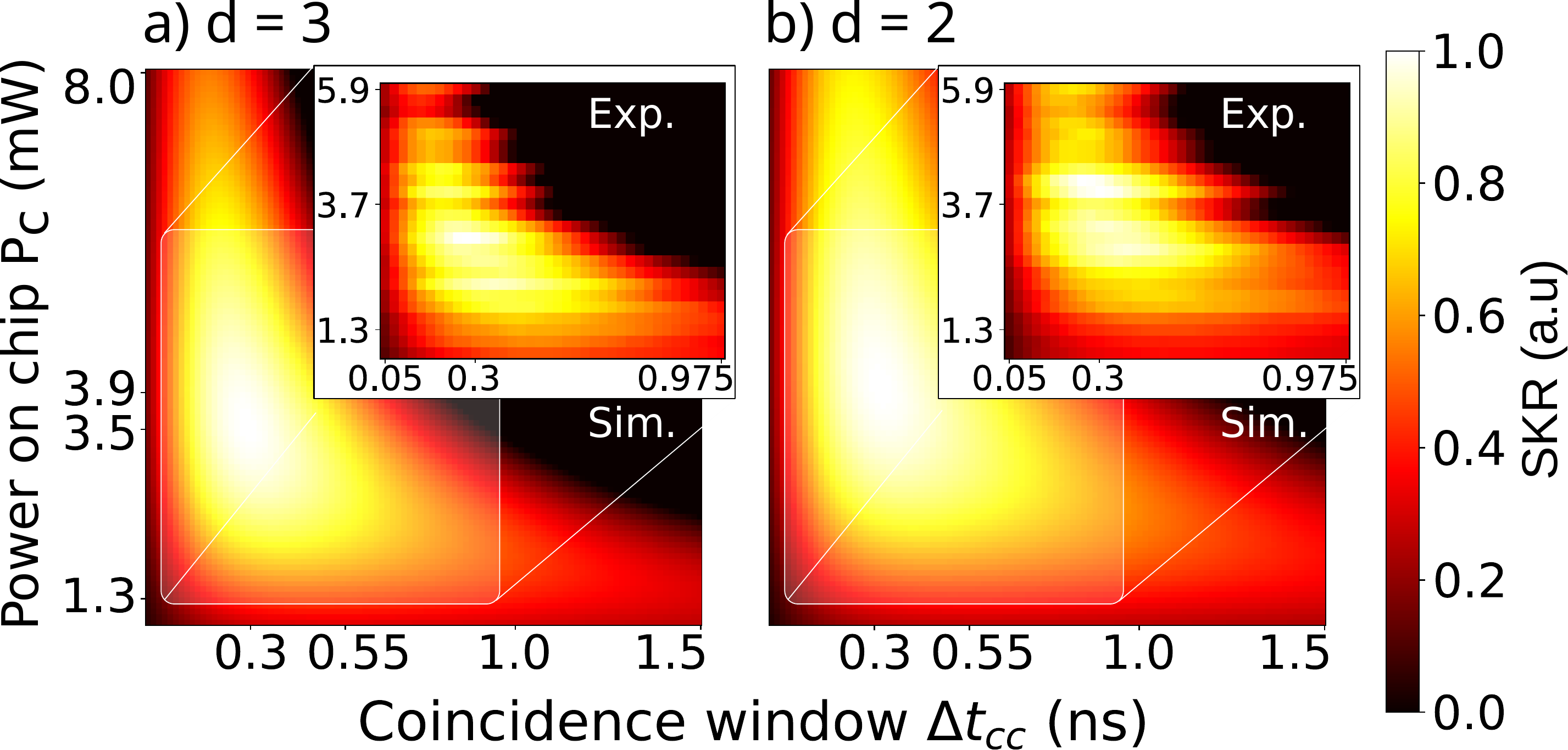}
    \caption{Simulated (Sim.) and experimental (Exp.) Secure Key Rate (SKR) as a function of coincidence window $\Delta t_{cc}$ and power on chip $P_c$ for $d=3$ qu$d$it in \textbf{a)} and for $d=2$ qu$d$its in \textbf{b)}.
    The experimental $\text{SKR}(\Delta t_{cc},P_c)$  represent the highlighted smaller area of the simulations.
    The optimal power on chip and coincidence window are $P_{op}^{3\text{D}}=3.5$ mW and $\Delta t_{op} ^{3\text{D}} = 285$ ps for $d$=3 qu$d$its and $P_{op}^{2\text{D}}=3.9$ mW and $\Delta t_{op} ^{2\text{D}} = 310$ ps for $d$=2 qu$d$its.
    }
    \label{Fig.Cartography}
\end{figure}
\begin{figure*}
    \centering
    \includegraphics[width=\linewidth]{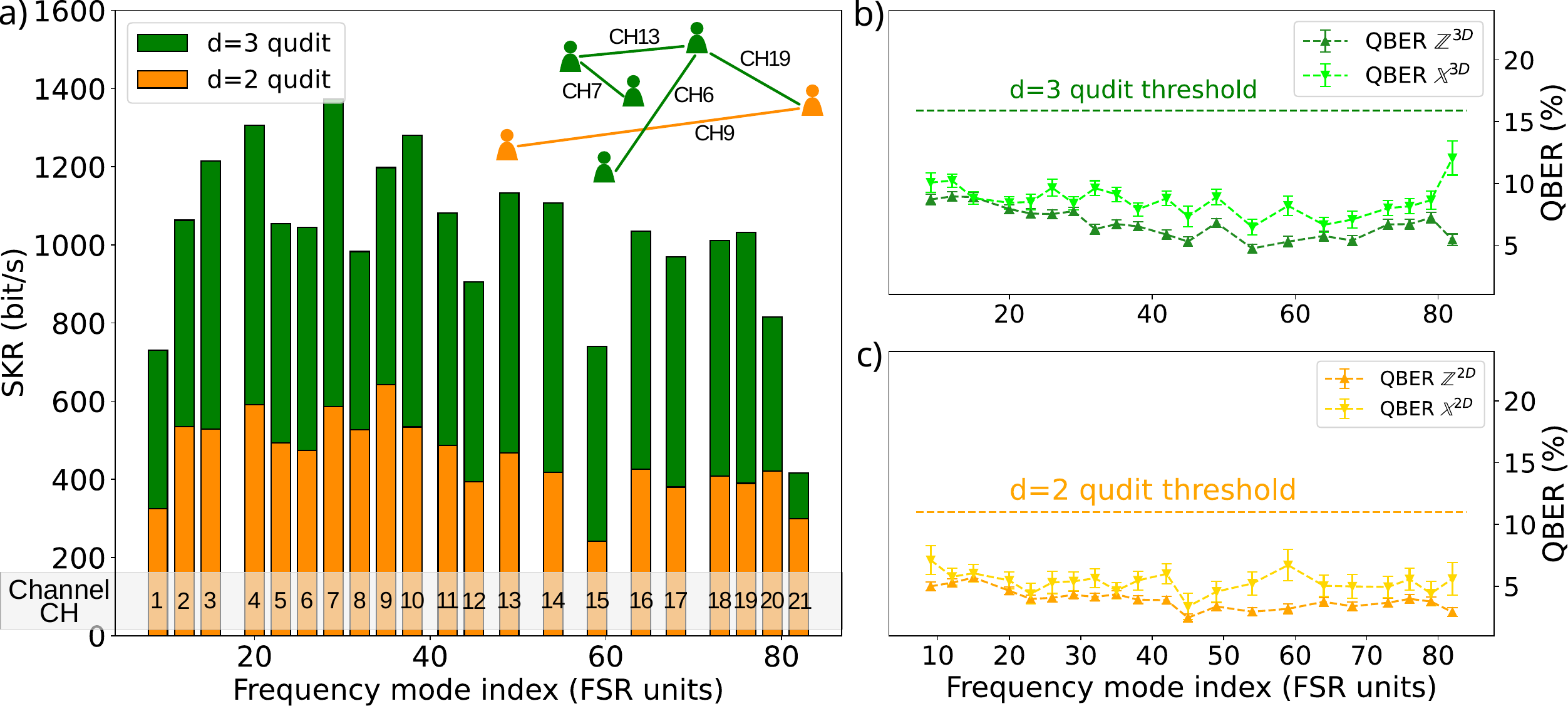}
    \caption{\textbf{a)} Experimental Secure Key Rate (SKR) of $21$ QKD channels at 0 dB applied attenuation. We manually select the QKD channels and exclude the frequency modes with coincidence count rates below $1$ kHz (Fig. \ref{fig:Setup}b). Each channel is $3$ resonance-wide ($63$ GHz) and indexed by the spectral separation from $\omega_p$ of the central frequency-mode, in FSR units.
    Each channel can be deployed with $d=3$ qu$d$its or with $d=2$ qu$d$its.
    An example of a multi-dimensional quantum network architecture that operates quantum channels with qutrits and qubits in parallel is depicted in the inset.
    The associated QBERs for measurements in $\mathbb{X}$ (light colored) and $\mathbb{Z}$ (dark colored) basis are shown in \textbf{b)} for $d=3$ and in \textbf{c)} for $d=2$ qu$d$it implementation.
    The horizontal dotted lines are the QBER thresholds for positive SKR, of $15.9\%$ and $11\%$ respectively \cite{cerf_security_2002}.
    }
    \label{Fig.Performances}
\end{figure*}
where $n$ is the index of the central frequency mode fixed for a given quantum channel.
We define \{$|\omega _l \rangle\}$ as the frequency-bin encoding for the logical state $\{| l \rangle \}$, with $l \in \{0,...,d-1\}$ of the natural basis $\mathbb{Z}$, whether it is for a signal or an idler photon.
Alice and Bob actively and independently decide to measure their photon in the natural basis $\mathbb{Z} = \{|z_k\rangle = |\omega_k\rangle\}$ or in the superposition basis $\mathbb{X} = \{|x_k\rangle= 1/\sqrt{d}  \sum_{l}^{d-1} e^{2i\pi k l/d} |\omega_l\rangle \}$, where $k \in \{0,...,d-1\}$, and $d$ is qu$d$it state dimension.
With this notation, the qubit measurement bases are given by
\begin{table}
    \centering
    \begin{tabular}{ccc}
    $\mathbb{Z}^{2\text{D}}$ basis:& &$\mathbb{X}^{2\text{D}}$ basis:  \\
    
    $|z_0\rangle= |\omega_0\rangle$\ & & $|x_0\rangle = \frac{1}{\sqrt{2}}\left(|\omega_0\rangle+|\omega_1\rangle \right)$ \\
    $|z_1\rangle= |\omega_1\rangle$\ & & $|x_1\rangle = \frac{1}{\sqrt{2}}\left(|\omega_0\rangle-|\omega_1\rangle \right)$, \\

\end{tabular}
\end{table}

and the qutrit measurement bases are given by
\begin{center}
\begin{tabular}{ccc}
    $\mathbb{Z}^{3\text{D}}$ basis: & & \ $\mathbb{X}^{3\text{D}}$ basis: \\
    $|z_0\rangle= |\omega_0\rangle$\ & & $|x_0\rangle = \frac{1}{\sqrt{3}}\left(|\omega_0\rangle+|\omega_1\rangle +|\omega_2\rangle \right)$ \\
    $|z_1\rangle= |\omega_1\rangle$\ & & $|x_1\rangle = \frac{1}{\sqrt{3}}\left(|\omega_0\rangle+e^{2i\pi/3}|\omega_1\rangle +e^{4i\pi/3}|\omega_2\rangle \right)$ \\
    $|z_2\rangle= |\omega_2\rangle$\ & & $|x_2\rangle = \frac{1}{\sqrt{3}}\left(|\omega_0\rangle+e^{4i\pi/3}|\omega_1\rangle +e^{8i\pi/3}|\omega_2\rangle\right)$ \\
\end{tabular}
\end{center}

An example of the frequency-bin qu$d$it $d=2$ BBM92 QKD implementation is shown in Fig. \ref{fig:Setup}c, where the Bell state $|\Psi\rangle^{2\text{D}}_n$ is shared between Alice and Bob.
Alice measures the signal photon in the $\mathbb{Z}^{2\text{D}}$ basis, where the Electro-Optic-Modulator (EOM) is turned off, and the input frequency modes $|S_{n-1}\rangle$ (which corresponds to $|\omega_0\rangle_A$) and $S_{n+1}$ ( which corresponds to $|\omega_1\rangle_A$) are routed to different single photon detectors, effectively performing the $\langle S_{n-1}|\Psi\rangle_n^{2\text{D}}$ and $\langle S_{n+1}|\Psi\rangle_n^{2\text{D}}$ projections. 

Bob projects the idler photon in the $\mathbb{X}^{2\text{D}}$ basis. PF$_1$ attributes a phase of $0$ to $|I_{n-1}\rangle$ ($|\omega_0\rangle_B$) and a phase $\varphi$ to $I_{n+1}$ ($|\omega_1\rangle_B$). The intensity of $|I_{n-1}\rangle$ and $|I_{n+1}\rangle$ are also equalized to ensure the normalization factor.
The EOM is driven by a single tone Radio Frequency (RF) signal at  frequency $\Omega_\text{rf} = \text{FSR}$, and mixes the frequency modes in the common frequency channel $I_n$ isolated by PF$_2$, effectively performing the $\frac{1}{\sqrt 2}(\langle I_{n-1}| +e^{i\varphi} \langle I_{n+1}|)|\Psi\rangle_n^{2\text{D}}$ projection, where $\varphi \in \{0, \pi \}$, which correspond to $\langle x_0|\Psi\rangle_n^{2\text{D}}$ and $\langle x_1|\Psi\rangle_n^{2\text{D}}$ projections \cite{kues_-chip_2017,lukens_frequency-encoded_2017}.

Analogous to $d=2$, any $d$ qu$d$it frequency-bin encoded QKD implementation can use the same PF-EOM-PF hardware configuration, given the EOM can mix all the frequency-bins on one common frequency channel.
In the $d=3$ qu$d$it case, the Bell state $|\Psi\rangle_n^{3\text{D}}$ is distributed to Alice and Bob.
The $\mathbb{X}^{3\text{D}}$ basis now consists of the $\frac{1}{\sqrt3}(\langle I_{n-1}| + e^{i\varphi} \langle I_{n}|+e^{2i\varphi} \langle I_{n+1}|)|\Psi\rangle_n^{3\text{D}}$ projections, where $\varphi \in \{0, 2\pi/3, 4\pi/3 \}$, corresponding to $\langle x_0|\Psi\rangle_n^{3\text{D}}$, $\langle x_1|\Psi\rangle_n^{3\text{D}}$ and $\langle x_2|\Psi\rangle_n^{3\text{D}}$ projections \cite{lu_quantum_2019}.
Note that this configuration allows for the projection of a photon onto only one state of the superposition basis at a given time. 
By converting frequency-bin encoding to time-bin encoding, it becomes possible to measure the entire superposition basis simultaneously, as has been recently demonstrated with qubit systems in \cite{khodadad_kashi_frequency-bin-encoded_2025,tagliavacche_frequency-bin_2025}.

In the BBM92 QKD protocol, the rounds in which Alice and Bob measure different bases are discarded, therefore we will only take into account the case where the two users measured the same basis.
The total amount of coincidences for each projection is denoted as $C_{M}^{a,b}$, where the index $M \in \{ \mathbb{Z}, \mathbb{X}\}$ is the basis chosen by Alice ($a$) and Bob ($b$), and the exponents $a,b \in \{ 0,..,d-1 \}^2$ stand for the outcome of their measurement.
Therefore true coincidence counts are defined when $a=b$, and accidentals counts are defined when $a\neq b$.
The Quantum Bit Error Rate (QBER) represents the bit error rate in the key generated by Alice and Bob after basis reconciliation, and before error correction.
It is given by the ratio between accidental coincidence counts and the total amount of coincidence counts. 
The general expression of the QBER for a $d$-dimensional state is given by
\begin{equation}
\label{eqn:QBER}
\epsilon_M^{dD} = \frac{\sum_{a=0, b=0, a \neq b}^{d-1} C_{M}^{ab}}{\sum_{a=0, b=0}^{d-1} C_{M}^{ab}}.
\end{equation}
The Secure Key Rate (SKR) represents a lower bound of the secure information exchange rate between the two users, and is defined under infinite key length approximation by \cite{cerf_security_2002,sheridan_security_2010}
\begin{equation}
\label{eqn:SKR}
\text{SKR}^{dD} \geq \frac{1}{2}R_{\text{raw}}^{d\text{D}} \left[ \log_2(d) - fH_d(\epsilon_Z) - H_d(\epsilon_X)  \right],
\end{equation}
where $d$ is the qu$d$it dimension.
The factor $1/2$ comes from the sifting ratio in the bases reconciliation step, $R_{\text{raw}}^{dD}$ is the average coincidence rate between the two bases
\begin{equation}
\label{eqn:Rraw}
R_{\text{raw}}^{dD} = \sum_{i=0,j=0}^{d-1} \left(C_\mathbb{Z}^{ij} + C_\mathbb{X}^{ij} \right) /2\tau,
\end{equation}
with $\tau$ the integration time, and $H_d(x)$ is the generalized binary entropy function for d-level systems, given by \cite{sheridan_security_2010}
\begin{equation}
H_d(x) =-x \log((x/d-1)) - (1-x)\log(1-x).
\end{equation}
The term $fH_d(\epsilon_Z)$ represents the bits revealed during the error correction process, that can no longer be used in the secure key, where $f \geq 1$ is the post processing efficiency.
In our case, $f=1.2$ both for qubit and qutrit implementations.
Finally, the term $H_d(\epsilon_X)$ represents the fraction of the key lost after privacy amplification, required to nullify any potential information obtained by an eavesdropper, Eve.

\section{QKD link optimization}\label{section:QKD link optimization}

In this section we optimize the source and signal processing parameters of the QKD channel link composed of the resonances $14-15-16$, which is the highest coincidence rate triplet of the JSI Fig. \ref{fig:Setup}a, and maximize the SKR of the channel.
We simulate the SKR obtained for powers on chip $P_c$ ranging from $1$ to $8$ mW and for coincidence windows (acceptance time of the detectors) $\Delta t_{cc}$ ranging from $30$ ps to $1.5$ ns in Fig. \ref{Fig.Cartography}a for $d=3$ and b for $d=2$ qu$d$its. 
The simulation is based on the entanglement-based QKD model with a continuous-wave pumped Spontaneous Parametric Down Convertion (SPDC) source without cavity from \cite{neumann_model_2021}, adapted to SFWM in cavity. More details are given in Methods section \ref{methods}.
The smaller boxed regions for $P_c \in [1,6]$ mW for coincidence windows $\Delta t_{cc} \in [50,975]$ ps have been measured experimentally, and the results are given in the insets. 
At such pumping powers, the multi-photon pair emission rate (which increases the accidental counts), and the two photon absorption \cite{Yin07} (which limits the photon pair generation rate) are significant, but taken into account in the SKR calculation.
The optimal experimentally obtained parameters are shown in Tab. \ref{tab:OptimParamsTab}.

\begin{table}{}
\centering

\begin{tabular}{*{5}{w{c}{0.145\linewidth}|}w{c}{0.145\linewidth}}

  qudit  & $P_{op}$ & $\Delta t_{op}$  & $\epsilon_{\mathbb{X}}$& $\epsilon_{\mathbb{Z}}$ & $g^{(2)}_h(0)$ \\ \hline
 d=3      & $3.5$ mW & $285$ps  & $8.1 \%$ & $7.7\%$ & $6.4\%$ \\ \hline
 d=2      & $3.9$ mW & $310$ps  & $5.3 \%$ &  $4.4\%$ & $7.7\%$ \\ \hline
\end{tabular}
\caption{Values of the optimized QKD parameters for d=3 and d=2 qudits. $P_{op}$= optimal power on chip, $\Delta t_{op}$= optimal coincidence time window, $\epsilon_{\mathbb{X}}$ and $\epsilon_{\mathbb{Z}}$ represent the QBERs of the $\mathbb{X}$ and $\mathbb{Z}$ basis respectively, associated with the highest SKR for optimal power and coincidence time window and $g^{(2)}_{h}(0)$= heralded second order correlation function at optimal power and coincidence window.}
\label{tab:OptimParamsTab}
\end{table}

\begin{figure*}
    \centering
    \includegraphics[width=0.9\linewidth]{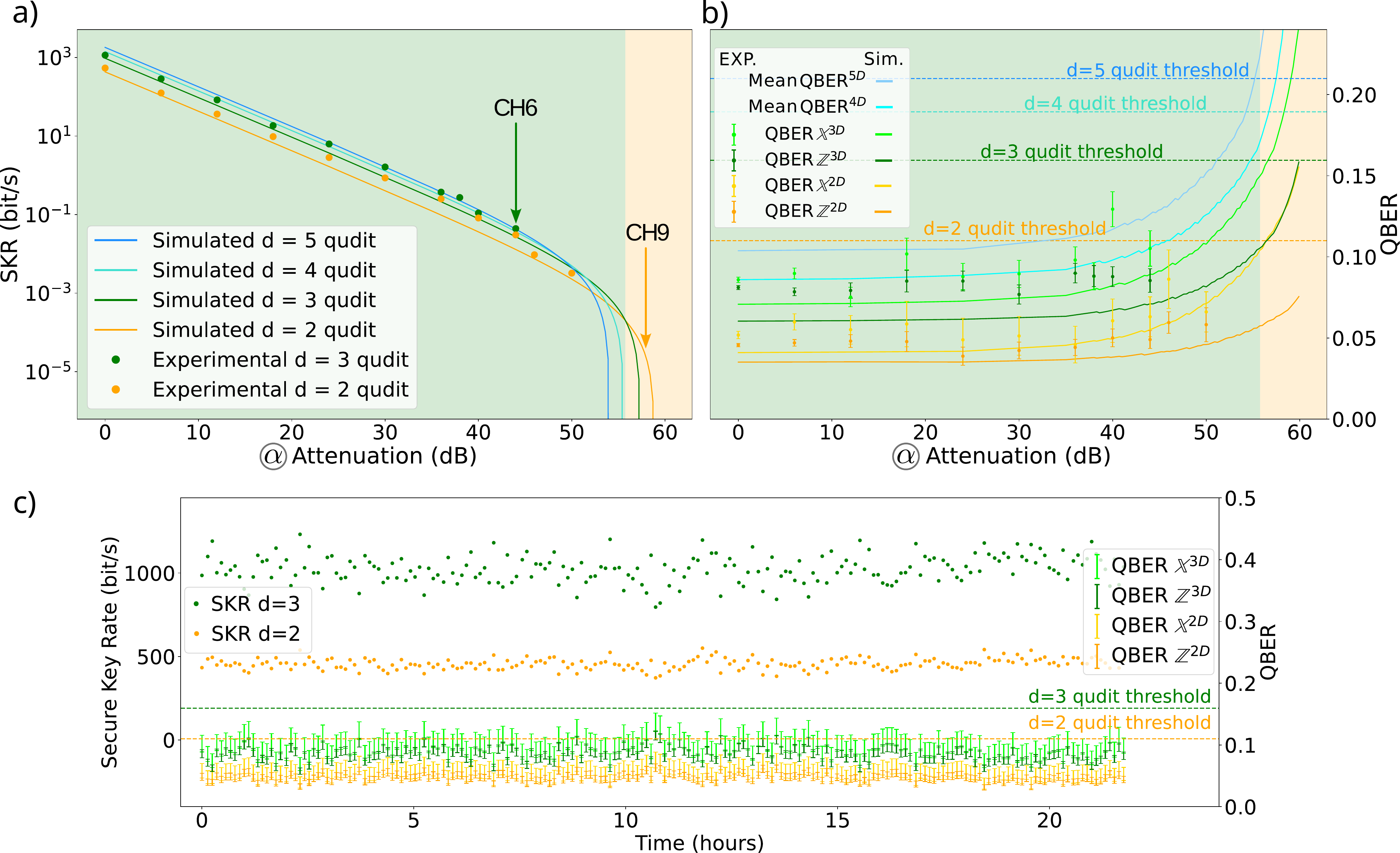}
    \caption{Experimental (dots) for $d=2$ and $d=3$ and simulated (plain line) for $d=2$ to $d=5$ \textbf{a)} Secure Key Rate (SKR) and \textbf{b)} Quantum Bit Error Rate (QBER) scaling with total attenuation $\alpha$ on channel CH6. 
    The green background indicates the region where $\text{SKR}^{3\text{D}} > \text{SKR}^{2\text{D}}$ up to $55$ dB of attenuation ($275$ km), and the orange background indicates $\text{SKR}^{2\text{D}} > \text{SKR}^{3\text{D}}$ from $55$ to $59$ dB attenuation ($295$ km). For example, the shorter quantum channel CH6 can deploy $3$-dimensional states, while the longer quantum channel CH9 can operate with $2$-dimensional states in parallel, using the same PF-EOM-PF configuration as in Fig. \ref{fig:Setup}.
    The horizontal dashed lines of pannel b correspond to the QBER threshold below which a positive key can be extracted. 
    \textbf{c)} SKR and QBER every $500$ s, over $21$ hours, for $d=3$ and $d=2$ qu$d$it dimension, using only the active frequency stabilization feedback loop.
    }
    \label{Fig.Distance}
\end{figure*}

The optimal powers on chip $P_{op}$ found in Fig. \ref{Fig.Cartography} are $P_{op}^{3\text{D}} = 3.5$ mW for $d=3$ and $P_{op}^{2\text{D}} = 3.9$ mW for $d=2$ qu$d$it implementations.
Higher dimensional qu$d$its are more affected by the QBER than the qubits, therefore the optimal pumping power decreases as the qudit dimension increases. 
Indeed, for an error rate $\epsilon$, the binary entropy function $H_d(\epsilon)$ increases with the dimension $d$.
This effect is compensated by the higher error tolerance of higher dimensional qudits. 
The QBER threshold, below which no secure key rate can be generated, is $15.9\%$ for $d=3$ and $11\%$ for $d=2$, assuming $\text{SKR}^{d\text{D}}(f=1)=0$ \cite{cerf_security_2002}.

The optimal coincidence window for $d=3$ and $d=2$ qudit implementations are $\Delta t_{op}^{3\text{D}} = 285$ ps and  $\Delta t_{op}^{2\text{D}} = 310$ ps, respectively. 
It is mainly influenced by the temporal length of the photons, of $410$ ps at half maximum, which does not depend on the dimension chosen. 
Nevertheless, the number of orthogonal projections $C_{M}^{a\neq b}$ in equation (\ref{eqn:QBER}) increases with $d(d-1)$, while the number of colinear projections $C_{M}^{ab}$ increases with $d$, which leads to a higher accidental over total counts ratio for higher $d$-dimensional states. 
For instance, $\epsilon_{\mathbb{Z}}^{3\text{D}}$ is around $1.8$ times higher than $\epsilon_{\mathbb{Z}}^{2\text{D}}$ for the same signal to noise ratio.
The lower optimal coincidence window compensates for this effect.
Moreover, the heralded second order auto-correlation function at optimal parameters equals $g^{(2)}_{h}(0)^{3\text{D}} = 6.4\%$ for $d=3$ and $g^{(2)}_{h}(0)^{2\text{D}} = 7.7\%$ for $d=2$ qu$d$its.
The $g^{(2)}_{h}(0)$ value does not constitute a security risk, as the increase in QBER due to multi-photon pair emissions in absence of an eavesdropper outweighs the potential information gain of an eavesdropper from such events.

In entanglement-based QKD networks, the QBER alone suffices to assess the security of the quantum channel, as it encapsulates all relevant information about potential eavesdropping and channel disturbances.

\section{Frequency domain quantum network}

We estimate the secure key rate for $d=3$ and $d=2$ qu$d$it based quantum channels, and demonstrate a high dimensional QKD network in the frequency domain. We also perform distance scaling and stability measurements of the secure key rate.

We perform the BBM92 protocol using the Bell states in eq. (\ref{eqn:Psi3}) and (\ref{eqn:Psi2}) for the available frequency modes in Fig. \ref{fig:Setup}b at $0$ dB applied attenuation $\alpha$.
We manually select $21$ $63$-GHz-wide ($3$ FSRs) to constitute a QKD network, excluding the frequency modes below $1$ kHz of coincidence count rate.
We set the power on chip $P_c$ to the optimal power $P_{op}^{3\text{D}}$ and $P_{op}^{2\text{D}}$ found in Fig. \ref{Fig.Cartography} for $d=3$ and $d=2$ qu$d$its, respectively. The coincidence window $\Delta t_{cc}$ is selected in post-processing to maximize the SKR for each mode and attenuation $\alpha$.

We measure an average SKR of $1024$ bit/s across the $21$ QKD channels with a maximum of $1374$ bit/s for the channel CH6 for $d=3$ qu$d$its, and an average of $456$ bit/s with a maximum of $642$ bit/s for the channel CH9 for $d=2$ qu$d$its. The results are shown in Fig. \ref{Fig.Performances}a. We give the associated QBERs in Fig. \ref{Fig.Performances}b and c for $d=3$ and $d=2$ qu$d$its respectively.
The average QBERs over both $\mathbb{X}$ and $\mathbb{Z}$ bases are $8.4\%$ for $d=3$ and $4.7\%$ for $d=2$, well below the positive key rate threshold plotted horizontally of $15.9\%$ and $11\%$ \cite{cerf_security_2002} respectively.

We perform distance scaling measurements of the channel CH6, applying a total attenuation $\alpha$ symmetrically ($\alpha/2$ on each path using PF$_1$ in Fig. \ref{fig:Setup}).
The channel CH6 has been selected for its SKR being representative of the average SKR across the $21$ quantum channels.

We measure the SKR$^{d\text{D}}$ and the QBER$^{d\text{D}}$ for $d=3$ and $d=2$ qu$d$it implementations as functions of the applied attenuation $\alpha$ and indicate the results with dots in Fig. \ref{Fig.Distance}a and b, respectively.
The plain lines correspond to the simulated SKR$^{d\text{D}}$ and QBER$^{d\text{D}}$ for $d=2$ to $d=5$ qu$d$it implementations, estimated at optimal parameters $P_{op}^{d\text{D}}$ and $\Delta t_{op}^{d\text{D}}$ for each $d$ and $\alpha$. A dark count rate of $350$ Hz per detector has been chosen for the simulated SKRs and QBERs.
The colored background serves as visual guide, marking the threshold at $55$ dB (corresponding to $275$ km, assuming $0.2$ dB/km), beyond which the SKR of $d=3$ qudit implementation drops below that of the $d=2$ qu$d$it implementation.
The maximum communication range is $59$ dB or $295$ km considering optical losses of $0.2$ dB/km.

Both the SKR and the QBER increase with the dimension $d$, whereas the communication range decreases as $d$ increases. 
A strategy for high dimensional QKD networks is to use high dimensional states for short quantum channels, and progressively decrease the dimension for longer quantum channels.
Indeed, the number of required projections per basis scales with the quantum states dimension, which enhances the per-photon information capacity via a larger encoding alphabet, but also reduces the Signal to Noise Ratio (SNR), due to increased dark count rates at long channel distances for high dimensional states with respect to qubits.
The flexibility of frequency-bin encoding enables quantum network users to benefit from both the increased SKR of the $d>2$ qu$d$its, up to $55$ dB and from the from the higher SNR of qubits, up to $59$ dB.
We also pushed our BBM92 protocol implementation with $5$-dimensional frequency-bin encoded states up to $18$ dB of total attenuation to CH6. This results in a secure key rate SKR$^{5\text{D}} \approx 300$ bit/s and QBERs of $\epsilon_{\mathbb{Z}}=10.6\%$ and $\epsilon_{\mathbb{X}}=21.4\%$ as the limited modulation index $\mu \leq 1.2$ of our EOM limits the efficient projections of $5$-dimensional states in the superposition basis $\mathbb{X}$.

Measurements of the SKR and the QBER over time in Fig. \ref{Fig.Distance}c show that the frequency stabilization of our system enables stable and continuous communication over more than $21$ hours.
The increased standard deviation of $d=3$ ($82$ bit/s) with respect to $d=2$ qu$d$its ($32$ bit/s) comes from the precision of PF$_1$ to apply the phase $\varphi$ and from its fluctuations, which affects more the higher dimensional states.
After $\approx 36$h the resonance frequency shifts out of the stabilization feedback loop range.
To regain a stable regime, the re-initialization of the frequency stabilization feeback loop takes about $90$ seconds, and the automatised-alignment of the fibers takes about $3$ minutes.
This demonstrates autonomous QKD infrastructure that requires little to no surveillance.
In more realistic scenarios, the projections on the $\mathbb{X}$ basis would require an active phase stabilisation, as implemented in \cite{tagliavacche_frequency-bin_2025}, to compensate for the phase shifts due to thermal and mechanical fluctuations of the environment.

\section{Discussion}

At the moment of writing, three other frequency-bin encoded BBM92 QKD implementations have been reported \cite{henry_nm_2024,khodadad_kashi_frequency-bin-encoded_2025,tagliavacche_frequency-bin_2025}, harnessing qubits only.
To the best of our knowledge this work is the first frequency encoded demonstration with 3-dimensional states.
Our reconfigurable frequency-bin encoding QKD network enables the coexistence of $21$ QKD channels of $d=2$ and $d=3$ - qu$d$it encodings using the same hardware.
This work builds on our QKD network implementation with 12 channels in \cite{henry_nm_2024}, but as the frequency-modes are spatially separated in this work, we do not use any guard-band to separate the QKD channels, increasing the number of available channels for qu$d$its.
We also tune the source and detection parameters to optimize the Signal to Noise Ratio (SNR) and increase the SKR and the communication range for qubits and qutrits.

The recent implementations with qubits ($d=2$) \cite{khodadad_kashi_frequency-bin-encoded_2025,tagliavacche_frequency-bin_2025} use a conversion from frequency-bin encoding to time-bin encoding, enabling the simultaneous measurement of the projections in superposition basis. Our PF-EOM-PF configuration allows for the measurement of only one projection in the superposition basis at a time. The approach in \cite{khodadad_kashi_frequency-bin-encoded_2025,tagliavacche_frequency-bin_2025} is nonetheless compatible with the $d=2$ and $d=3$ scalable implementation described here.

The \textbf{SKR of our frequency-bin encoding QKD} implementations is modest when compared to state-of-art implementations in polarization \cite{li_high-rate_2023,chen_integrated_2021} and time-bin \cite{terhaar_ultrafast_2023} encoding achieving $10^5-10^7$ bit/s, but it can be increased by reducing the losses in the system.
The loss budget from the on-chip photon pair generation to the detectors sums up to $17.5$ dB per user ($\geq 20 $ dB for $\mathbb{X}$ basis) in the current setup, cf. Methods section \ref{setup}.
The losses between the grating couplers and the fibers, currently of $3.95$ dB, can be reduced to less than $0.7$ dB using sub-wavelength index grating couplers techniques \cite{Benedikovic15}.
Integrated versions of the pump rejection filter NF on Fig. \ref{fig:Setup}a have recently been demonstrated with insertion losses below $0.1$ dB \cite{Oser2019}. On-chip integration of PFs  \cite{cohen_silicon_2024,wu_-chip_2024} and EOMs \cite{hu_-chip_2021} would reduce even further the insertion losses, with the potential to increase the SKR by at least 2 orders of magnitude.

We achieve a competitive \textbf{communication range} above $250$ km, alongside polarization encoding implementations \cite{li_high-rate_2023,wengerowsky_entanglement-based_2018,appas_flexible_2021} and frequency-bin encoding implementations \cite{khodadad_kashi_frequency-bin-encoded_2025}, while time-bin encoding implementations achieve $150$ km \cite{terhaar_ultrafast_2023,yu_quantum_2025}.
We extend the communication range to $295$ km by optimizing the on chip power and the coincidence window to reduce the signal to noise ratio.
For comparison, the communication range achieved by a satellite-to-ground BBM92 QKD implementation is $2000$ km \cite{chen_integrated_2021}.

\begin{figure}
    \centering
    \includegraphics[width=\linewidth]{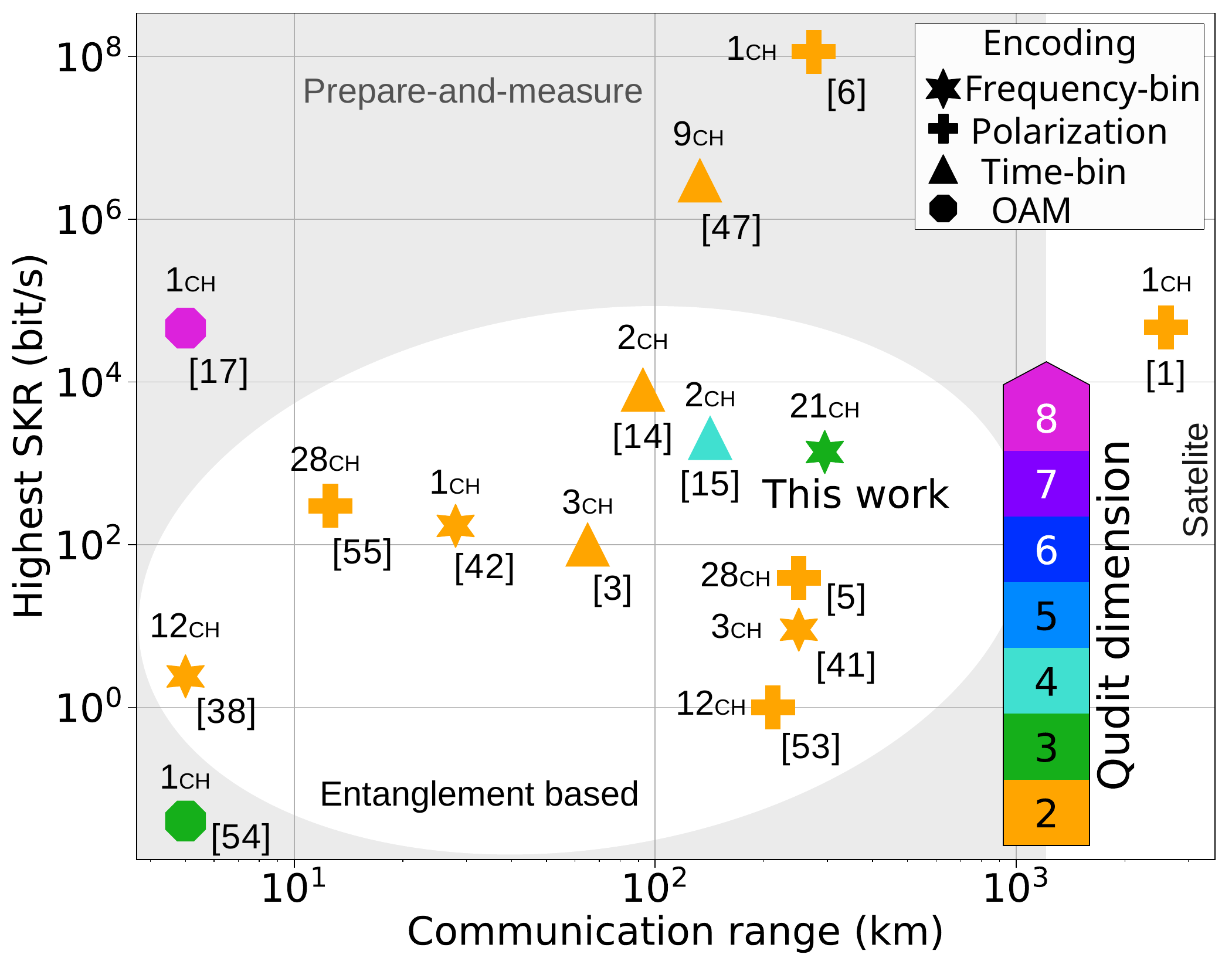}
    \caption{An overview of QKD network implementations, with respect to the highest achieved single channel SKR -estimated at $0$ km quantum channel length- and estimated communication range, at normalized length to $0.2$ dB/km.
    We highlight high dimensional implementations using a color code for the qu$d$it dimension. The degree of freedom of the encoding is indicated by the marker.
    The size of the QKD network is given as the number of quantum channels $\#_{CH}$.
    The gray area and the two white areas visualize prepare-and-measure based protocols, entanglement based and satellite-to-ground entanglement based implementations.
    The SKR and communication range of the satellite-to-ground implementations are not normalized.
    }
    \label{Fig.SOA}
\end{figure}

In pursuit of scaling to \textbf{higher dimensions}, we explored frequency-bin state encoding up to $d=5$.
Employing EOMs with higher modulation indices and bandwidths up to $119$ GHz \cite{hu_-chip_2021} and driving them simultaneously with multiple radiofrequency tones would enable the efficient generation of $d=12$ qu$d$its, further increasing the maximally achievable SKR. 
This approach preserves the efficiencies and communication ranges achieved lower $d$ qu$d$it.
Fig. \ref{Fig.SOA} contextualizes our work within the broader landscape of QKD implementations targeting for high-dimensional encoding schemes.
OAM encoding (octagon) gives versatile and cost-efficient access to high dimensional states \cite{groblacher_experimental_2006}, which enables SKRs higher than $40$ kbit/s with $d=4$  \cite{sit_high-dimensional_2017} and $d=8$ \cite{bouchard_experimental_2018} qu$d$its implementations. OAM encoding has nonetheless limited compatibility with standard optical fibers and distance scalability, achieving typically less than $2$ km of communication range (set to $5$ km in Fig. \ref{Fig.SOA}).
Time-bin encoding (triangle) is compatible with qu$d$its and optical fibered quantum networks \cite{yu_quantum_2025}.
A dispersive optics QKD protocol employed $32$ time slots for information encoding \cite{lee_large-alphabet_2019}. However in time-bin encoded QKD protocols the SKR and the dimensionality of the quantum states are not independent parameters, and stabilizing multi-arm interferometers remains challenging.

With respect to \textbf{network scalability}, The parallel operation of $2$ \cite{Steiner2023,yu_quantum_2025}, $3$ \cite{fanyuan_robust_2022,khodadad_kashi_frequency-bin-encoded_2025}, $9$ \cite{terhaar_ultrafast_2023} and $12$ \cite{henry_nm_2024,wengerowsky_entanglement-based_2018} QKD channels has been implemented in various degrees of freedom.
By leveraging the low FSR of our photon pair source and the inherent multiplexing capacity of frequency-bin encoding (star), we demonstrate a QKD network operating with 21 QKD channels, which is comparable to recent demonstrations up to 28 \cite{joshi_trusted_2020,appas_flexible_2021} QKD channels in polarization encoding (cross).
We can increase the number of QDK channels to $38$ by setting the spectral width of each channel to $42$ GHz ($2$ resonance-wide), supporting only 2-dimensional states implementations, instead of $63$ GHz (3 resonance-wide) that supports 2 and 3-dimensional states implementations.
The number of QKD channels can be further increased by lowering the FSR of the source and reducing the channel spectral width accordingly, and by using a wider bandwidth PF.

QKD networks have various assets depending on the implementation, and a wider picture would include metrics such as the deployment stage and real-time or post-processed measurements.
The prospects for estimated distance scaling of our implementation show frequency-bin encoding is a contender for QKD network implementations, as shown in Fig. \ref{Fig.SOA} with the chosen metrics of highest achieved SKR and communication range.

\section{Conclusions}

We have demonstrated an experimental proof of principle of a chip-based multi-user, multi-dimensional BBM92 QKD protocol using frequency-bin encoded qu$d$its of dimension $d=3$ and $d=2$, with possibility of further increasing $d$ using the same hardware configuration.
We used a $21$ GHz FSR silicon micro resonator enabling $80$ biphoton frequency modes in a $5$ THz bandwidh, supporting parallel operation of $21$ $3$-resonance-wide quantum channels.

We have optimized the signal to noise ratio of $d=3$ and $d=2$ qu$d$it QKD implementations to increase the secure key rate and extend the communication range.
We found an average secure key rate of $1024$ bit/s for $d=3$, and of $456$ bit/s for $d=2$ qu$d$its among the $21$ frequency-bin encoded QKD channels.

The extended communication range of our system provides a competitive advantage, while the secure key rate can be enhanced through the integration of filtering and measurement devices on-chip. 
We achieved a maximal SKR of $1374$ bit/s is with qutrits, and a communication range of $295$ km, in the asymptotic regime and symmetric attenuation scenario, with qubits.
We leverage the advantages offered by both qubit and higher-dimensional quantum states using a single reconfigurable hardware to address user-specific needs -offering higher transmission rates with high-dimensional states and extended communication ranges with two-dimensional states.

Even higher dimensional quantum states would require EOMs with higher modulation bandwidths, while a $5$-dimensional states protocol has partially been explored.
Our QKD network can support up to $38$ QKD channels with qubits, and lowering the FSR of the source would further increase the number of accessible QKD channels.
The system exhibits stable operation over 21 hours, but field deployment would require additional phase stability over long distance fibers.

Our proof of principle demonstration of frequency-bin encoded entanglement-based QKD network constitutes a step toward the development of scalable and robust multi-dimensional quantum networks and an evidence of the potential of frequency encoding and silicon based implementations.

\textbf{Acknowledgments}
This work has been supported by Region Ile-de-France in the framework of DIM SIRTEQ.
This work benefited from Sitqom ANR project no. ANR-SITQOM-15-CE24-0005, QuanTEdu-France no. ANR-22-CMAS-0001 France 2030, and QComTestbed no. ANR22-PETQ-0011 and was partially supported by the European Union's PHOQUSING project GA no. $899544$. The device fabrication was partially performed within the C2N technological platforms and partly supported by the RENATECH network and the General Council of Essonne.

\section{Methods}\label{methods}
\subsection{Experimental setup}\label{setup}
The pump setup of \ref{fig:Setup}a consists of a Tunics T100R LASER from Yenista. 
It has a basic wavelength tunable mode with a step size of $1$ pm, and a fine-tune mode with a step size of $0.1$ pm. 
The $0.5$ mW output power of the laser is amplified by an Erbium-Doped-Fiber-Amplifier (EDFA) set at $23$ dBm ($\approx200$ mW) and attenuated with a voltage controlled Thorlabs VOA, to reach the desired power on chip $P_c$.
A $30$ GHz-wide EXFO XTM-50 band pass filter centered around $1539.95$ nm is used to filter out the amplified spontaneous emission light of the laser.
The sample temperature $T_{\text{sample}}$ is actively stabilized at $26\pm0.1$ °C with a PID temperature controller, and the room temperature is set at $23\pm0.4$ °C with air conditioning. The device fabrication was partially performed within the C2N technological platforms and partly supported by the RENATECH network and the General Council of Essonne.

The pump and collection optical fibers are coupled to the bus waveguide of the micro resonator through grating couplers. The fibers are mounted on piezoelectric translation stages, enabling automated alignment. This configuration allows for an end-to-end optical transmission of $-7.89$ dB to be achieved within less than 3 minutes.
The collected light is routed into a notch filter (NF) with a bandwidth of $95$ GHz ( $0.75$ nm) and an optical rejection of $25$ dB. The rejected pump laser light is routed to a power meter, facilitating active frequency stabilization, described in the next Methods section \ref{Stabilization}, while single photons are routed to the measurement devices.
We use Programmable Filters (PF) from FINISAR (WaveShaper 4000A) that have a maximum optical rejection of $60$ dB. 
The total laser pump optical rejection of the system adds up to $145$ dB.
Lithium niobate Electro-Optic Modulators (EOM) from iXblue (now Exail) are driven by a single tone Radio Frequency (RF) signal at frequency $\Omega_{rf}=21.23$ GHz from a multiple-output phase locked ANAPICO Signal Generator.
The photons are detected using Superconducting Nanowire Single Photon Detectors (SNSPDs) from Single Quantum with efficiencies $\approx 76\%$ at telecom wavelength. 
We count the coincidences as two detection events that occured within a coincidence window of $\Delta t_{cc}$ using a TimeTagger 20 from Swabian Instruments.

We have a loss budget of $17.5$ dB per user, from the photon pair on chip generation generation. 
The chip-to-fiber coupling losses represent $3.95$ dB, which corresponds to half of the end-to-end transmission, assuming equal coupling efficiencies for the pump and the collection fibers.
The insertion losses of the NF, PF, and EOM are respectively of $1.5$, $4$ and $3$ dB.
For the projections in the $\mathbb{X}$ basis has an additional $3$ dB of losses due to the EOM creating side bands outside the computational space.

\subsection{Stabilization of the system} \label{Stabilization}

\begin{figure}
    \centering
    \includegraphics[width=\linewidth]{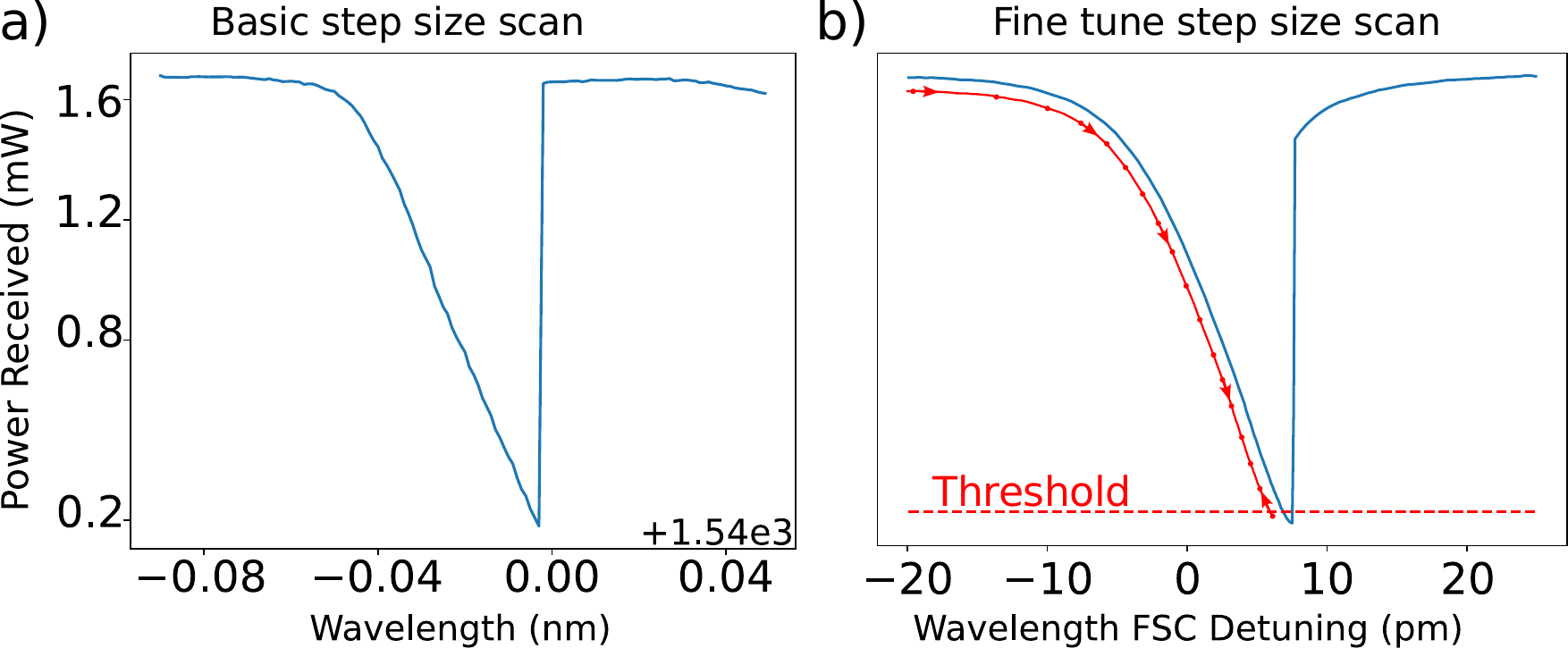}
    \caption{\textbf{a)} Power meter received power, filtered out by the notch filter versus the basic scan of the pump frequency, with steps of $1$ pm, around the resonance mode near $1540$ nm.
    \textbf{a)} Power meter received power versus the Fine Scan (FSC) of the pump frequency, with steps of $0.1$ pm.
    The red horizontal dashed line shows the threshold above which the active frequency stabilization script will maintain the power.
    The red dotted line is an eye-guide on the part of the resonance that the active frequency stabilization tries to keep the frequency in.
    }
    \label{Fig.frequency}
\end{figure}

The pump laser filtered out by the NF is measured by a power meter which enabling active frequency stabilization by adjusting the pump frequency to minimize the optical transmission through the bus waveguide of the resonator. 
Examples of the active frequency stabilization scans are showed in Fig. \ref{Fig.frequency} 
During the initialization phase, we scan a firs time the spectral area near the desired resonance, which is shown in Fig. \ref{Fig.frequency}a to be at $1539.985$ nm. 
The resonance is asymmetric with a sharp edge due to thermal bistability at high power.
We measure a maximum power received $P_{max} = 1.6$ mW on the power meter, wihch corresponds to a power on chip of $P_c = P_{max} \times 10^{(3.95+1.46)/10} = 5.5$ mW, given by half of the end-to-end transmission and the insertion losses of the NF.
We perform a second, fine tune scan showed in Fig. \ref{Fig.frequency}b, during which we measure a minimum power received $P_{min}$ and define a threshold $T = P_{min} + 0.02 \times P_{max}$. 
During the stabilization phase, the power received $P_r$ is measured iteratively.
A positive detuning step of the pump frequency is then estimated, if $P_r > T$, or a negative detuning if $P_r < T$. 
The threshold indicates the safe spectral detuning from the sharp edge of the resonance, which would otherwise detune the system out of resonance.
This method enable stable positioning of the pump frequency in resonace with the microresonator, close to the minimum of transmission despite the central spectral position of the resonance shifting.

Moreover, we can automatically align the pump and the collection fibers using piezo controlled fiber mounts. 
Using the same setup, we detune the pump frequency off resonance, and vary the voltage applied by the piezo controllers, and hence the position of the pump and collection fibers, on $x$ and $y$ axis. 
We measure the power after each step and optimize the position the fibers to maximize the transmission of the bus waveguide of the source.
The automatic alignment of the fibers has not been used during any of the measurements presented in this work, but it can correct mechanical shifts during multiple day operation period of the source.

\subsection{Simulation}\label{simulation}

In the entanglement-based QKD with continuous-wave pumping demonstration of \cite{neumann_model_2021}, an SPDC source is modeled.
A theoretical framework for true and accidental coincidences is established, which scale linearly and quadratically with the brightness of the source, respectively.
The raw coincidence rate $R_{\text{raw}}^{2\text{D}}$ and the QBER $\epsilon^{2\text{D}}$ are then derived.
We adapt the model to SFWM in cavity. 
The arrival time histogram between two detetors, namely the $g^{(2)}$ correlation function between signal and idler photons, is fitted with a Voigt profile,
\begin{equation}
    V(x,\sigma,\gamma) = \int_{-\infty} ^{+\infty} G(x^\prime, \sigma) L(x-x^\prime, \gamma) dx^\prime,
\end{equation}
accounting for the Lorentzian $L$ resonance of the resonator with $\gamma = 99.3$, and for the Gaussian response of the single photon detectors with $\sigma = 123.2$.
In our model, the power on chip $P_c$ is too high to consider that the true coincidences from the photon pair generation rate and the background noise arising from multi-photon pair emissions have a linear and quadratic scaling with $P_c$, as considered in \cite{neumann_model_2021}.
Indeed, above $P_c \geq 500$ $mu$W, competing effects like two-photon absorption saturate the generation of photon pairs \cite{Yin07}.
The study of such effects is not the main purpose of this work, therefore we measure the $g^{(2)}$ correlation function for every experimental $P_c$ in Fig. \ref{Fig.Cartography}. 
We fit the true and accidental coincidences scaling with $P_c$.
To simulate the scaling with the coincidence window $\Delta t_{cc}$, we multiply the true coincidences by $\eta^{t_{cc}}$ determined by
\begin{equation}
    \eta^{\Delta t_{cc}} = \int_{-\Delta t_{cc}} ^{+\Delta t_{cc}} V(t_{cc}) dt_{cc},
\end{equation}
and the accidental coincidences by $2\Delta t_{cc}$.
The QBER and the SKR are then simulated using the equations (\ref{eqn:QBER}) and (\ref{eqn:SKR}) for a wide $P_c$ and $\Delta t_{cc}$ area, for $d=2$ to $d=5$.

\newpage

\textbf{References}
\bibliographystyle{apsrev4-2}
\bibliography{refs}

\end{document}